%% file: article_acm.tex
\documentclass[sigconf]{acmart}

\usepackage{amsmath,amsfonts}
\usepackage{algorithmic}
\usepackage{graphicx}
\usepackage{textcomp}

\usepackage{url}
\usepackage{framed}
\usepackage{tabularx}
\usepackage{booktabs}
\usepackage{comment}
\usepackage{float} 

\usepackage{changepage}
\usepackage{tcolorbox}
\usepackage{enumitem}

\newcommand{\highcell}[1]{{\color{highlight} \textbf{#1}}}

\definecolor{highlight}{HTML}{0f8aFF}

\def\BibTeX{{\rm B\kern-.05em{\sc i\kern-.025em b}\kern-.08em
    T\kern-.1667em\lower.7ex\hbox{E}\kern-.125emX}}

\usepackage[framemethod=TikZ]{mdframed}
\definecolor{backgroundpage}{HTML}{F8F5F4}

\newenvironment{coloredframe}[2][]{
    \mdfsetup{
        skipabove=5pt, skipbelow=5pt,
        innertopmargin=5pt, innerbottommargin=5pt,
        hidealllines=true, leftline=true,
        innerlinewidth=1pt, innerlinecolor=#2, 
        linewidth=0pt,
        backgroundcolor=#2!10
    }
    \begin{mdframed}}
    {\end{mdframed}}

\definecolor{interviewcolor}{HTML}{b8b0ff}

\newcommand{\interviewquote}[2]{
   \begin{coloredframe}{yellow}
    ``#1''. (#2)
    \end{coloredframe}
}

\AtBeginDocument{%
  \providecommand\BibTeX{{%
    \normalfont B\kern-0.5em{\scshape i\kern-0.25em b}\kern-0.8em\TeX}}}

\copyrightyear{2024}
\acmYear{2024}
\setcopyright{rightsretained}
\acmConference[ICSE-SEET '24]{46th International Conference on Software Engineering: : Software Engineering Education and Training}{April 14--20, 2024}{Lisbon, Portugal}
\acmBooktitle{46th International Conference on Software Engineering: : Software Engineering Education and Training (ICSE-SEET '24), April 14--20, 2024, Lisbon, Portugal}
\acmDOI{10.1145/3639474.3640063}
\acmISBN{979-8-4007-0498-7/24/04}

\begin{document}

\title{Building Collaborative Learning: Exploring Social Annotation in Introductory Programming}

 \author{Francisco Gomes de Oliveira Neto}
 \orcid{0000-0001-9226-5417}
 \affiliation{%
   \institution{Chalmers and the University of Gothenburg \\
    Dept. of Computer Science and Engineering}
   \city{Gothenburg}
   \country{Sweden}
 }
 \email{francisco.gomes@cse.gu.se}

 \author{Felix Dobslaw}
 \orcid{0000-0001-9372-3416}
 \affiliation{%
   \institution{Mid Sweden University\\
   Dept. of Quality Mngmt, Communication and Inf. Systems}
   \city{Östersund}
   \country{Sweden}
 }
 \email{felix.dobslaw@miun.se}

\renewcommand{\shortauthors}{de Oliveira Neto and Dobslaw}

\input{sections/00_abstract.tex}



\maketitle

\input{sections/01_introduction.tex}
\input{sections/02_background_rw.tex}

\input{sections/03_methodology.tex}
\input{sections/04_results.tex}
\input{sections/05_discussion.tex}
\input{sections/06_conclusions.tex}

\bibliographystyle{ACM-Reference-Format}
\bibliography{bibliography}

\end{document}

%% file: sections/00_abstract.tex
\begin{abstract}
The increasing demand for software engineering education presents learning challenges in courses due to the diverse range of topics that require practical applications, such as programming or software design, all of which are supported by group work and interaction. 
Social Annotation (SA) is an approach to teaching that can enhance collaborative learning among students. In SA, both students and teachers utilize platforms like Feedback Fruits, Perusall, and Diigo to collaboratively annotate and discuss course materials. This approach encourages students to share their thoughts and answers with their peers, fostering a more interactive learning environment. 
We share our experience of implementing social annotation via Perusall as a preparatory tool for lectures in an introductory programming course aimed at undergraduate students in Software Engineering. We report the impact of Perusall on the examination results of 112 students. Our results show that 81\% of students engaged in meaningful social annotation successfully passed the course. Notably, the proportion of students passing the exam tends to rise as they complete more Perusall assignments. In contrast, only 56\% of students who did not participate in Perusall discussions managed to pass the exam. We did not enforce mandatory Perusall participation in the course. Yet, the feedback from our course evaluation questionnaire reveals that most students ranked Perusall among their favorite components of the course and that their interest in the subject has increased.
\end{abstract}

%% file: sections/01_introduction.tex
\section{Introduction}

Programming is one of the first topics taught in many engineering disciplines. Large classes of students typically have their first contact with programming when teachers explain basic programming constructs as statements and are shown examples of output produced by executing code~\cite{xie2019theory,nelson2017comprehension}. Most of that knowledge is not introduced to students before higher-level education. Moreover, most students are unfamiliar with the technological content knowledge associated with teaching and learning programming (e.g., development environments, installation of compilers or interpreters). Exercising those skills already in the first class can easily overwhelm students, particularly those without any prior knowledge of programming~\cite{xie2019theory}. Besides the content itself, students must learn how to explain their algorithms to each other, i.e., explain to peers the steps that they followed to solve a specific problem~\cite{qian2017students,whalley2006australasian}. This is particularly challenging when students need to collaborate towards a solution in, e.g., a project course. 

Teaching approaches focused on flipping the classroom or performing active learning have shown effective results in improving the exam results of students~\cite{gren2020flipped,freeman2014active}. Particularly, preparing for lectures by reading material or watching videos has been one of the main tools used to allow teachers and students to focus their time together on solving problems and discussing different solutions to the problems. However, those studies did not investigate the collaborative dimension of students working and learning together. 

Social Annotation (SA) is a pedagogical approach that fosters collaborative learning among students, enabling them to jointly engage with course materials, discuss concepts, solve problems, and compare their annotations with peers~\cite{crouch2001peer,mazur1997peer}. Recent research highlights the effectiveness of collaborative learning in computer science and programming education, with students showing increased engagement and improved learning outcomes~\cite{lee2013can,porter2011peer}. SA has garnered overwhelmingly positive student responses, underscoring its potential for enhancing the educational experience~\cite{gao2013case}.

Social Annotation requires an online platform facilitating communication and knowledge sharing among students as they interact with course resources, such as textbooks, exercises, and video lectures. Feedback Fruits, Diigo, and Perusall are a few examples of such tools. For instance, in Perusall~\footnote{\url{https://www.perusall.com/}}, instructors share course materials, allowing students to asynchronously and collaboratively generate annotations by highlighting specific sections within the material --- whether those annotations target timestamps in videos, sentences or paragraphs in text, or sections of a web page. These annotations serve as a medium for students to write their understanding, identify challenging concepts, and seek clarification through questions or comments.

This interactive annotation process encourages students to articulate their comprehension of concepts and pinpoint difficulties. These annotations trigger discussions among students, as other students provide their own explanations, hence fostering collaborative learning dynamics within the environment. Teachers can use such discussions to discover why certain ideas remain unclear to students and cover them in discussions with the entire class.

Our goal is to investigate the impact of social annotations in Perusall when teaching programming to first-year students in the Software Engineering and Management bachelor program at the University of Gothenburg (Sweden). Particularly, we aim to verify whether students sharing their understanding of the course material with their peers affects their performance in the course exam. We compare the results of 112 students who used Perusall to prepare for each lecture by completing reading assignments. Particularly, our report targets the following research questions:

\begin{coloredframe}{teal}
\noindent \textbf{RQ1: Do students of a programming course engage in non-compulsory social annotation activities?}\\
\textit{Yes.} Most students (on average 78\% of 112 students) engaged in social annotations throughout all 18 course lectures. However, roughly 20\% of those students created meaningful comments that showed their understanding of the topic.\\

\noindent \textbf{RQ2: Does social annotation engagement have an impact on the students' grades and passing rates?}\\
\textit{Yes.} Students who engaged in social annotation by creating more meaningful comments had, proportionally, better grades and passing rates than those who were less engaged in social annotations.
\end{coloredframe}

Our results reveal that students who create meaningful comments in Perusall and engage in discussion with their peers have better grades in the course. Failure rates decrease for those students who use Perusall more in the course. Moreover, the course feedback questionnaire reveals a positive response from students using Perusall in the course, which aligns with existing findings in literature~\cite{lee2013can,novak2012educational}. On the other hand, the course feedback indicates that many of the students were not motivated to engage in Perusall discussions.

This paper is structured as follows. Section~\ref{sec:rw} presents related work regarding teaching programming, social annotation, and previous studies with Perusall. Section~\ref{sec:course} provides context to our investigation by sharing course details, such as student population and course structure.\footnote{Some details were omitted due to the double-blind review process. Some of the course artefacts (feedback form, example of the exam, course material can be shared after the reviewing process.)} We detail the data collection and analysis in Section~\ref{sec:method}. We present results and findings from our research questions in Section~\ref{sec:results}, followed by a discussion involving student feedback, lessons learned and the limitations of our report (Section~\ref{sec:discussion}). Lastly, we conclude and outline future work in Section~\ref{sec:conclusion}.

%% file: sections/02_background_rw.tex
\section{Related Work}
\label{sec:rw}


Learning programming goes beyond the skill of writing code according to a syntax (theory), it also requires students to trace the execution by predicting outputs and state changes \cite{lopez2008tracing,nelson2017comprehension}, as well as explaining the code to other programmers \cite{whalley2006australasian}. Those skills are connected but distinct from one another, such that instruction models aim to refine them. For instance, the Theory of Instruction model proposed by Xie et al. ~\cite{xie2019theory} highlights four programming skills based on reading and writing code using knowledge at a machine level (semantics) or at task level (templates). Those four skills are progressively obtained by the student by reading semantics, writing semantics, reading templates, and writing templates. 

Our goal is not to propose or evaluate such models, rather we investigate the prospects of social annotations that can later be used in combination with such models to bring forward the aspect of metacognitive thinking. In Perusall, students are encouraged to share their cognitive processes of understanding when posing questions or providing answers within shared course materials~\cite{miller2018perusall}, and prompting students to engage in metacognitive thinking can enhance their abilities in reading and writing code~\cite{loksa2016selfregulation,loksa2022metacognition}. 

Perusall, or social annotation, has not been widely investigated in the context of teaching and learning programming yet. Meyer and M\"{u}ller found significant challenges in implementing Perusall in a Data Structures and Algorithms course~\cite{meyer2019SAinCS}. The primary difficulty was in maintaining student motivation to annotate materials and participate in online discussions. Similarly, we observed that many of our students failed to produce annotations that effectively demonstrated their understanding of the subject matter. These experiences underscore the necessity of promoting a shift towards a culture of continuous learning

Other subject areas have shown promising results in using SA. In a comparative study, Suhre et al. identified a positive correlation between active participation and examination results with Perusall in eight different courses in social sciences~\cite{suhre2019students}. They further found that engagement can be fostered through the tool if certain criteria are respected including stimulating texts and assignment formulations, appropriate group sizes, the a-priori providing of good annotation examples, as well as timely feedback from instructors. Those findings align with other studies focused on social annotation with other platforms, where researchers see improvements in learning engagement~\cite{hwang2007multimedia}, attention~\cite{huang2008annotations}, peer communication, and sense of community~\cite{kalir2020social}.

The papers above provide insights into diverse methods for evaluating student performance and learning. In this experience report, we establish correlations between Perusall activity and students' exam performance to discern distinctions among groups of students actively engaged in social annotation. Although exam scores offer a limited perspective on student performance~\cite{grissom2015using}, they have been employed in prior research that investigates Perusall and social annotation, enabling the exploration of student engagement and learning~\cite{huang2008annotations,lin2013harnessing,miller2018perusall}. This approach allows us to draw parallels with our findings. In future studies, we intend to explore additional dimensions of student performance, including their sense of belonging and learning progression throughout the course.


%% file: sections/03_methodology.tex
\section{Case Course: Context and Scope}
\label{sec:course}

\begin{figure*}[ht!]
    \centering
    \includegraphics[width=0.9\linewidth]{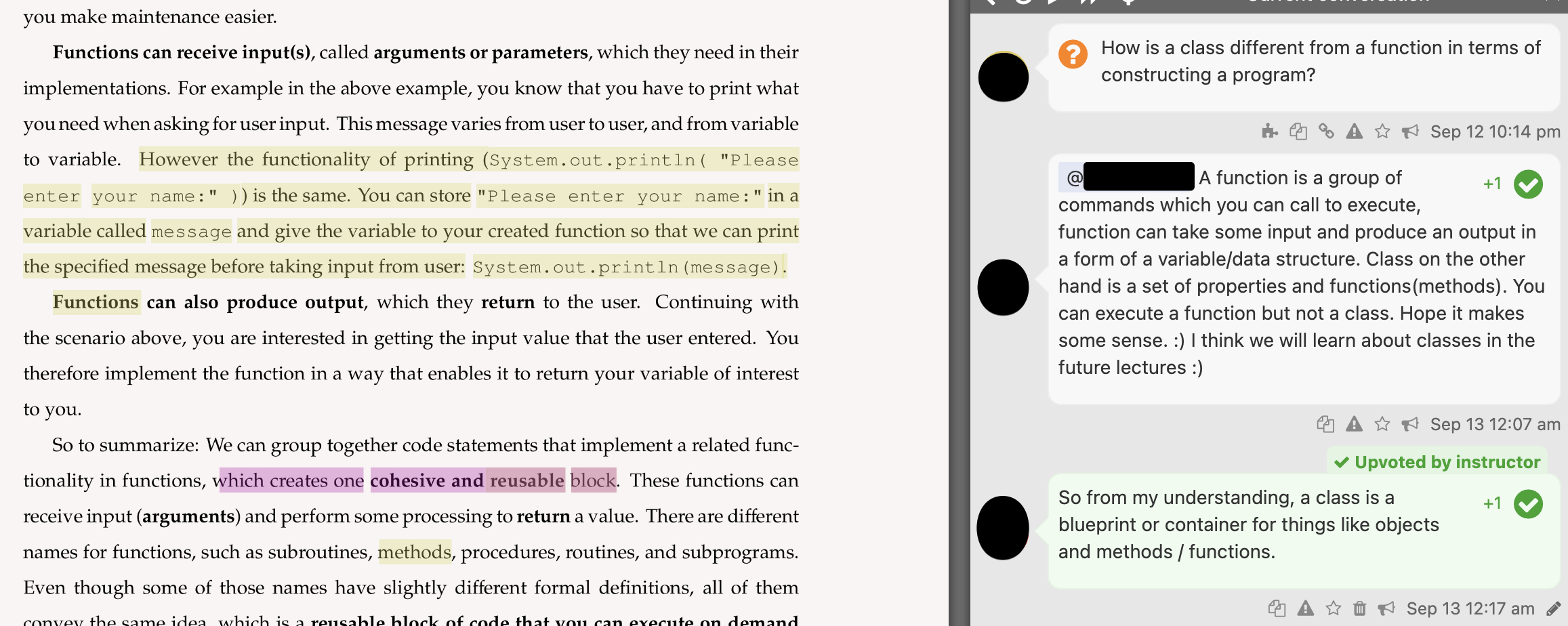}
    \caption{Example of student interaction in Perusall. Three students comment on the highlighted (pink) annotation in the course material about the lecture on Functions in Java. Students help each other understand the difference between reusable code for simple tasks (functions) and structural abstractions (classes). 
    }
    \label{fig:perusall-example}
\end{figure*}

We investigate the impact of Perusall in a course taught in the first study period of an international bachelor program in Software Engineering and Management at the University of Gothenburg (Sweden). The course is on Object-oriented Programming (OOP) and covers the following learning outcomes: (i) basics in procedural programming (e.g., printing, conditionals, loops, arrays and functions), and (ii) core concepts of OOP (e.g., classes, objects, encapsulation, polymorphism). The programming language taught in the course is Java.\\

\textbf{Course structure:} The course instance took place during 10 weeks in 2022 and had 143 registered students. Students were expected to dedicate 20 hours per week to the course, which would include time in lectures, laboratory sessions (focused on practical exercises), and self-studies at home. Students were offered three 2-hour lecture sessions, and three 2-hour lab sessions a week --- all of which \textit{non}-compulsory. For course completion, the students must submit: (i) three programming assignments done in groups of up to three; and (ii) a final individual written hall exam where students score between 0--100 points. The course was taught on campus by one course responsible, and eleven teaching assistants.\\

\textbf{Student background:} No entry requirements in programming or computer science applied, as this was the students' first programming course in the Program. Nonetheless, students may or may not have had previous programming knowledge (e.g., during their high-school education), leading to a heterogeneous sample of student backgrounds. Students took one other course in parallel in discrete mathematics with the same expected workload.\\

\textbf{Perusall and social annotation:} For each lecture, students were instructed to prepare by reading the material in the Perusall platform in the form of \textit{reading assignments}. Students complete these reading assignments by creating meaningful annotations, reading the material until the end, and engaging with the material (e.g., scrolling, highlighting text, etc.). The reading assignments are the main component of Perusall that fosters collaborative interaction between students before the lecture. Therefore, the completion of reading assignments will be our main metric to measure the social annotation element in the course.

To motivate students to complete reading assignments, the course instructor offered bonus points to students. A maximum of eight bonus points towards the exam were given to students creating \textit{meaningful annotations} or engaging in discussion in the Perusall material for the eighteen lectures.\footnote{Students were also informed that bonus points could not be used to cross the passing threshold of 50 points, i.e., a student could not pass the course through bonus points.} For each lecture, a 0.5 bonus point could be obtained. Thus, the maximum amount of points was achieved completing any 16 of the 18 Perusall tasks.

We used Perusall's definition of meaningful annotations and shared examples with the students at course start.\footnote{\url{https://support.perusall.com/hc/en-us/articles/360034824694-How-is-annotation-quality-defined-in-Perusall-}} In short, meaningful annotations are comments or questions that showcase the student's comprehension of the concepts discussed in the material. We exemplify a meaningful student exchange in Perusall through Figure \ref{fig:perusall-example} where three students discuss what a Function is. All three students received the bonus for that lecture as they made multiple similar comments throughout that lecture's material.

To cope with the large number of students, we used Perusall's algorithm that automatically assess the quality of annotations of students based on the content quality, the number of students' replies, length of text, among other features extracted from the annotation~\cite{king2019instructional}. The course instructor chose the holistic scoring strategy defined in Perusall which considers the annotations created, and whether the student has read the entire material before the lecture.\footnote{\url{https://www.perusall.com/hubfs/downloads/scoring-details.pdf}} The course instructor also determined that students needed to create at least two meaningful annotations to complete the reading assignment to foster discussion threads and communication between groups of students. The practical exercises and student-teacher interactions happened mainly during lectures. In case a student disagrees with Perusall's automatic score, the course instructor can revise and override Perusall's decision. Fine-tuning Perusall's accuracy is beyond the scope of our study, therefore, we acknowledge and discuss some limitations associated with its automated grading system in our threats to validity. \\ 

\textbf{Lecture format:} All lectures were hosted on campus. An average of 80 students showed up to class (55\% attendance rate). Each two-hour lecture was divided into two parts with a 10-min break in between. Part one was a Mentimeter\footnote{\url{https://www.mentimeter.com/}} session with multiple-choice questions regarding the Perusall material. We chose Mentimeter for its simplicity and to allow for anonymity. Students were told that the quiz participation had no influence on the grade and was optional. For each question, the answer statistics were presented live to the students, and the teacher initiated a discussion about the student's reasoning, particularly when the answers were not converging to the correct option. The second part of the lecture focused on applying the lecture topics with the help of one or two coding exercises solved together with the class.\\

\textbf{Written hall exam:} Students did a four-hour written exam with various questions focusing on tracing code, writing small classes or functions, and explaining the application and trade-offs of topics covered in the course. Students received between 0--100 points based on the quality of their answers. We applied the four-level grading scale below. The exams were anonymised by the examination office and graded by the course responsible. 

\begin{itemize}
    \item \textbf{Fail (U):} Assigned to students that scored less than 50 points in the exam.
    \item \textbf{Pass (3):} Given to students that scored between 50 and 69 points.
    \item \textbf{Pass with merit (4):} Given to students that scored between 70 and 84 points.
    \item \textbf{Pass with distinction (5):} This is the highest grade in the scale and is given to students that received points greater than or equal to 85.
\end{itemize}

\textbf{Course evaluation:} At the course's outset, five students volunteered to become student representatives who are the contact point for all students when the student collective wants to offer feedback regarding the teaching and learning throughout the course. Nonetheless, all students have direct channels to communicate with the course instructor. In the last week of lectures, \textit{all} students receive a questionnaire following the SEEQ feedback template~\cite{marsh1982seeq}. The questionnaire is closed before the written exam to reduce the risk of bias introduced by the examination experience. The course responsible and student representatives meet on two occasions: the first time halfway into the course to reflect on the course status and the teaching methods for possible intervention; the second meeting was a retrospective with the presence of the program manager and study administrators where the SEEQ questionnaire results were discussed. The information collected from those instruments helped the instructor to understand: (i) some of the main obstacles in the usage of Perusall, (ii) the frequency and level of satisfaction from students engaged in peer instruction, (iii) students' reactions to the teaching methods, and (iv) the self-reported impact on their learning. \\

\textbf{Prior Knowledge in Programming:} One of the main challenges in teaching first-year programming courses is the variance among students regarding their prior knowledge of programming. On one hand, having prior knowledge can lower students' motivation to engage in discussion about topics they are already familiar with. On the other hand, those students can also share their experiences with novice students to help them learn. Typically, programming is not taught in primary or secondary school, even though reality might change, given the benefits of introducing students earlier to programming~\cite{saeli2011teaching,xie2019theory}. 
The instructor estimated the prior knowledge of students by sending them an anonymous questionnaire with various programming-related questions before the first lecture (each question had an option ``\textit{I do not know\slash I cannot answer yet}''. From the sample of 115 respondents, 34\% of students could not answer what a String is, and 67\% of students did not know what an if-statement is. Both topics are basic programming constructs taught in the first week of the course. Therefore, we argue that prior programming knowledge is not a prevalent factor influencing our analysis in this instance of the course.

\section{Research Methodology}
\label{sec:method}

To prevent an unfair teaching environment and the risk of favoring or disadvantaging a particular group of students, we chose \textit{not} to conduct a controlled experiment. Instead, we gave students the choice by making social annotation an optional part of the course. Therefore, to answer our research questions, we used the individual results of the Perusall reading assignments together with the student's exam results. The feedback from the course evaluation questionnaire is used to discuss the qualitative aspects of the student's feedback about using social annotation. We refer to Perusall activity as the \textit{outcome} of the reading assignments made by each student. Each lecture had a corresponding reading assignment. There were three outcomes for each reading assignment:

\begin{itemize}
    \item \textbf{Skipped:} The student did not create a single annotation for that reading assignment, or they did not even read the material in Perusall.
    \item \textbf{Incomplete:} The student created at least one annotation in the material, but the content was not assessed as meaningful by Perusall's algorithm, i.e., the annotation did not convey the student's understanding of the subject covered.
    \item \textbf{Completed:} The student made at least two comments on annotations that were classified as meaningful according to Perusall's algorithms. These comments can be questions they asked, answers provided to other students or comments in discussion threads.
\end{itemize}

We compared those different types of activities in relation to the students' exam results (both points and grade). We analysed the exam results \textit{without} adding the bonus points from Perusall since this would otherwise introduce a bias towards passing students. We analysed the results of 112 students considering the intersection between those registering for Perusall during the course instance and taking the written exam.\footnote{Students who did not complete previous course instances also take the exam. Similarly, a subset of students participated in the course but decided to skip the written exam.} We anonymised the data set by removing all identity information from the records used throughout our analysis. 

For simplicity, the plots discussed in our results include IDs for each lecture. Table~\ref{tab:topics} maps each lecture ID to the corresponding subject covered by the reading material. For the remainder of the paper, we use the term assignments to refer to the \textit{reading assignments} in Perusall. Throughout our discussions, we consider that a student who completed a reading assignment conveyed their understanding of the topic to other students, which is one of the main goals of social annotation.

\begin{table}
\centering
\caption{List of topics covered in the course.}
\label{tab:topics}
\begin{tabular}{ll}
\toprule
  \textbf{ID}  & \textbf{Topic of the lecture:} \\
\midrule
  L01 & Variables, Types and Expressions \\
  L02 & User Input \\
  L03 & Conditionals \\
  L04 & Loops \\
  L05 & Arrays \\
  L06 & Basics in OOP \\
  L07 & Reference variables \\
  L08 & Encapsulation and immutable objects \\
  L09 & Collections - Lists, Sets and Maps \\
  L11 & Composition and Aggregation \\
  L12 & Inheritance \\
  L13 & Polymorphism \\
  L14 & Abstract Classes \\
  L15 & Exceptions and Error Handling \\
  L16 & Interfaces in Java \\
  L17 & OOP Design Principles: SOLID \\
  L18 & Files \\
  \bottomrule
\end{tabular}
\end{table}

\subsection{Scientific Ethics and Data Availability}

The University of Gothenburg is a State University under Sweden's principle of publicity (in Swedish, offentlighetsprincipen) which ensures transparency with the population.\footnote{\url{https://medarbetarportalen.gu.se/service-support/for-arbetsgivare/8.universitetet-ar-en-statlig-myndighet/}, available in Swedish.} Therefore, the public can request all exams (digital or printed) via a transparency office at the University. Students are made aware of such principles when admitted to the University. Nonetheless, we anonymise the data shared in this paper. To comply with scientific ethical guidelines, we also asked students and teachers for consent to use their course data (e.g., annotations, Perusall login data, exam results) during the first week of the course. We clarified that students could opt out of the study at any moment.

We share the files and scripts relevant to this experience report in our analysis package in Zenodo~\cite{replication2023perusall}.\footnote{\url{https://doi.org/10.5281/zenodo.10483184}} The CSV files include the exam points, grades and the classification of Perusall's annotation per student and lecture. We also share the report of the course evaluation.

%% file: sections/04_results.tex
\section{Results}
\label{sec:results}

Table \ref{tab:exam-sum} presents the results of the exam without any association to the Perusall activity. This allowed us to understand how the class performed. Based on the exam grade distribution, 45\% of the students failed the written exam. The one course parallel to this one showed a smaller yet similar failing rate (roughly 30\%). Below, we analyse our research questions by relating those percentages to the Perusall activity of students.

\begin{table}[h]
    \centering
    \caption{Distribution of exam grades. Students who failed the exam received a grade of U. Passing students received grades 3, 4 or 5 (highest grade).}
    \label{tab:exam-sum}
    \begin{tabularx}{\linewidth}{Xrrrr}
    \toprule
    \textbf{Grades:} &\textbf{U} & \textbf{3} & \textbf{4} & \textbf{5} \\
    \midrule
    \textbf{Number of students:} & 51 & 37 & 17 & 7 \\
    \textbf{Percentage of students:} & 45.5\% & 33.0\% & 15.2\% & 6.2\% \\
    \bottomrule
    \end{tabularx}
\end{table}


\subsection{RQ1: Level of Engagement in Social Annotation}

\begin{figure}
    \centering
    \includegraphics[width=\linewidth]{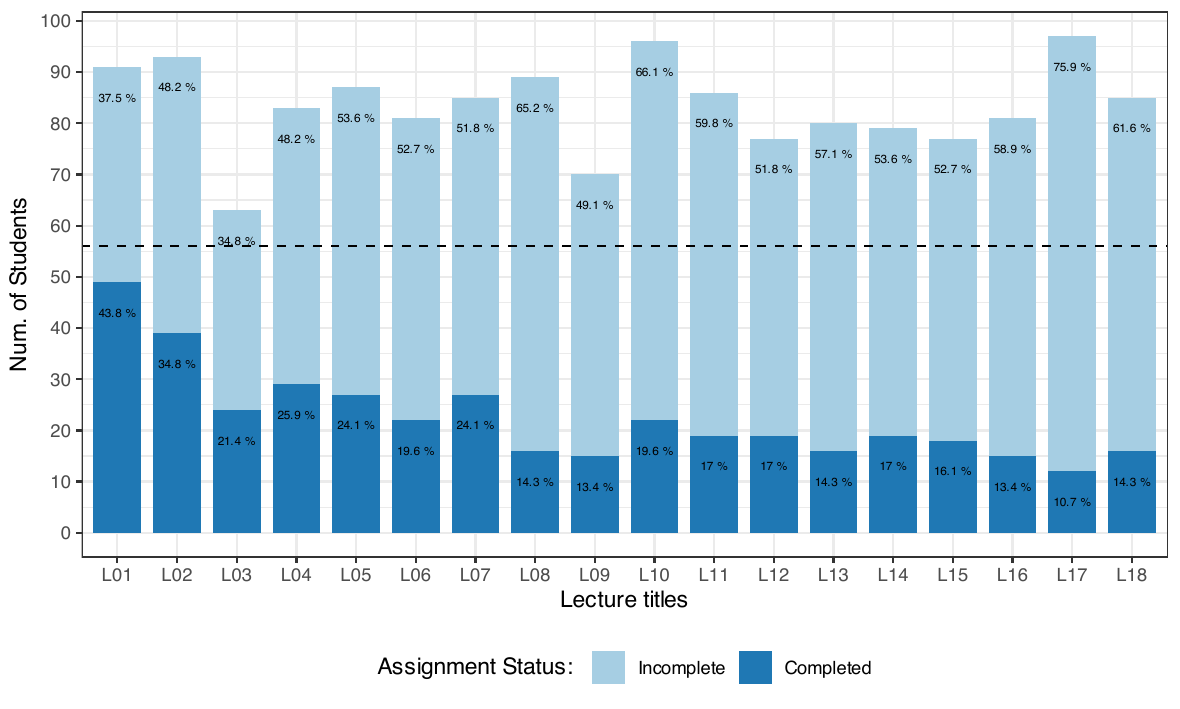}
    \caption{The overall distribution of social annotations with non-compulsory peer instruction. The dashed line intercepts the y-axis at half of the number of students (n = 56 students).}
    \label{fig:perusall-status}
\end{figure}

Figure \ref{fig:perusall-status} contains an overview of students' annotation patterns per reading assignment. More than 50\% of students participated in social annotation (Incomplete + Completed) with variations depending on the topic. An average of 78\% attempted or completed the interactions with the material. The lecture on loops (L03) had the least social annotation (44\% of the students skipped it), whereas L17 (OOP Design) had the most engagement (only 13\% skipped it). While the majority of the engaged students did not complete the reading assignments, students used the material throughout the course with no consistent signs of decline.

On the other hand, there was a decline in the proportion of students completing the reading assignments, hence indicating that fewer were making meaningful annotations\slash comments in Perusall. The drop was higher after L03 (loops) from 44\% to 21\% which is then sustained in different topics. The completion rate stayed consistently below 20\% after L08 (Encapsulation) and the other core OOP concepts. Note that the drop in completion rate did not affect the drop in reading assignments engagement as the proportion of incomplete assignments varies roughly between 40--60\% throughout all lectures. Below, we illustrate two contrasting annotations from different students about overriding methods in Lecture 12 (Inheritance).

\interviewquote{not very clear what does it mean functionality?}{Classified as Incomplete.}

\interviewquote{Why is [overriding] risky? I get that it could become a problem if a subclass needs to override a method, but doesn't. Is there any risk if everything works even if the subclass does not override the method and the superclass method is executed instead?}{Classified as Completed.}

The first comment was classified as incomplete because the student is not clear whether they mean the functionality of a shown piece of code or \textit{how} Inheritance and overridden methods work. In contrast, the student that completed the assignment conveys their current understanding of overriding risks (\textit{``I get that it could...''}) and their struggle to realise different risky scenarios (\textit{``... even if the subclass does not override...''}).

The lower percentage of completed assignments can be attributed to a variety of reasons. For instance, the learning curve to create meaningful annotations, lack of time to dedicate to social annotations, or a lack of accuracy in Perusall's automated algorithm to detect meaningful comments related to programming. Determining an accurate decline in the quality of annotations requires a more extensive and manual qualitative analysis of the text written by students which is outside the scope of our experience report. Therefore, we summarise our RQ1 findings below.

\begin{coloredframe}{teal}
\textbf{RQ1:} More than 50\% of students took part in the non-compulsory social annotation activities in Perusall. However, the percentage of students making meaningful comments or annotations decreased over time.
\end{coloredframe}

\subsection{RQ2: Impact of Social Annotations on Grades and Passing Rates}

\begin{figure}
    \centering
    \includegraphics[width=\linewidth]{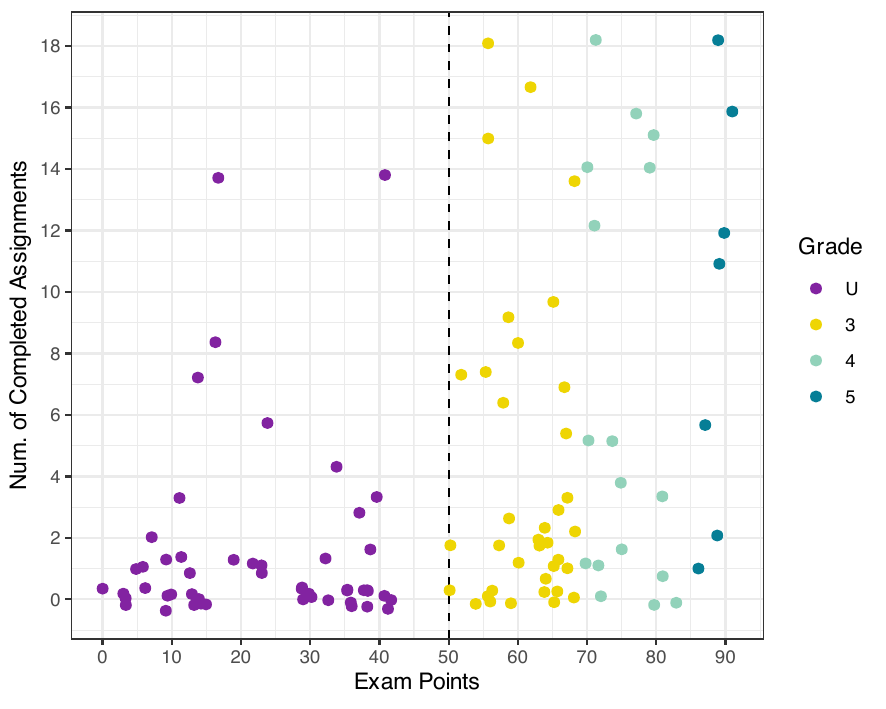}
    \caption{Exam points in correlation to the number of completed Perusall assignments. The dashed vertical line denotes the passing threshold (50 points).}
    \label{fig:individual-points}
\end{figure}

We measure social annotation based on the number of completed reading assignments. The reason behind our choice is that Perusall's algorithm mainly grades students based on the quality of their annotations which is one of the affecting factors in social annotation~\cite{huang2008annotations,lin2013harnessing}. Therefore, we assumed that students completing assignments have provided more insights to help their peers learn about the subject. Figure \ref{fig:individual-points} shows the correlation between the points obtained in the exam and the number of completed reading assignments. For students with none\slash little annotations, i.e., few completed assignments, no grade impact could be observed. As we increase the number of completed assignments, note that the number of students who failed started to decrease. Nonetheless, there were still many students who passed the exam without completing many reading assignments.

\begin{figure}
    \centering
    \includegraphics[width=0.9\linewidth]{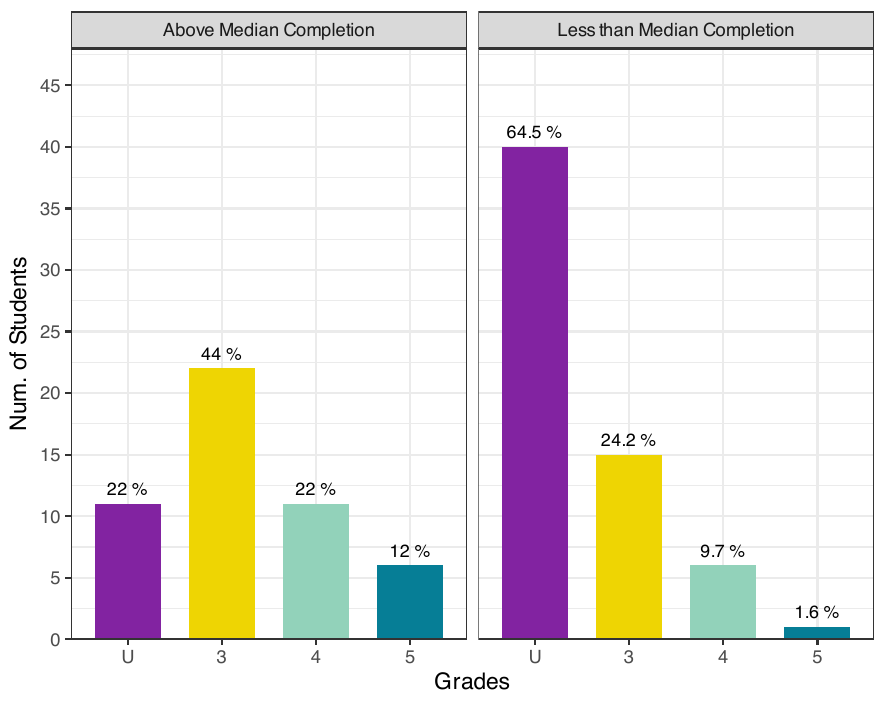}
    \caption{Percentage of students per each grade based on the expected number of completed assignments from passing students (more than 2 assignments). Students who completed above the median have better grades compared to those who do not.}
    \label{fig:divided-per-completed}
\end{figure}

To verify whether the number of completed assignments affected the exam results, we divided our sample into two groups and compared the proportion of students for each grade. We chose the median number of completed assignments for all passing students to divide the groups. Our reasons were two-fold: (i) the median conveys the expected number of completed assignments to pass the exam; (ii) the median divides the sample into two student groups of roughly equal size such that a similar proportion of students in all grades indicates that students would pass\slash fail the exam independently of their Perusall interactions. The median number of reading assignments completed for the passing students was 2, resulting in the two student cohorts in Figure \ref{fig:divided-per-completed}.

Of the 62 students who completed less than two reading assignments, 64.5\% (40) failed the exam. This proportion was almost three times higher than the proportion of failing students who completed at least two assignments (22\%). Moreover, the proportion of students in all passing grades is significantly higher in the group of students that completed the expected number of assignments to pass, particularly for the better grades 4 (6 vs. 11) and 5 (1 vs. 6).

\begin{figure*}
    \centering
    \includegraphics[width=0.9\linewidth]{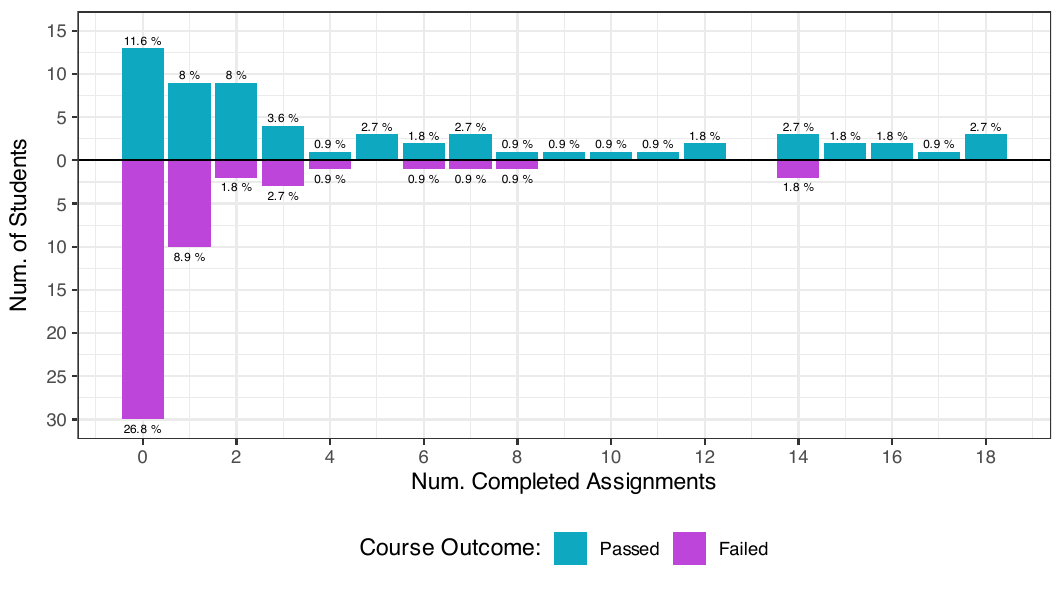}
    \caption{Course passing statistics per lecture. Completing many of the assignments (x-axis) has a large impact on passing the course, while little activity results in a high risk of failing. The ``negative'' y-axis is used simply to emphasise the difference between students who passed and those who failed.}
    \label{fig:diff-progression}
\end{figure*}

\begin{coloredframe}{teal}
\textbf{RQ2.1:} We observed a distinct grade distribution among students who actively participated in social annotation by completing the required number of assignments for passing. Notably, we identified a positive correlation between engagement and the distribution of exam scores and final grades. As students increased their involvement in social annotations within Perusall, we noted a decrease in the number of exam failures. Furthermore, a significant trend emerged, revealing that the majority of students who completed fewer than two assignments ended up failing the exam, while those who completed at least two assignments exhibited proportionally better performance in their exam grades.
\end{coloredframe}

Figure \ref{fig:diff-progression} contrasts the proportion of passing and failing students based on their \textit{corresponding} number of completed assignments. We see that the largest proportion of failing students are those who did not complete a single assignment (x = 0), which aligns with our observations above. Moreover, when completing more than 4 assignments the cumulative number of students passing the exam (25) is much higher than the number of students failing (5). Focusing on the middle range of completed assignments (4--9), few students fail and many more pass in that range. For the students highly engaged in social annotation (above 9 lectures) the correlation is even more apparent as only 2 (out of 19) students failed the exam.

\begin{coloredframe}{teal}
    \textbf{RQ2.2:} The majority of students who failed did not complete any reading assignments. After completing more than 4 assignments, the proportion of students passing the exam (22.5\%) is much higher than those that failed (4.5\%).
\end{coloredframe}

%% file: sections/05_discussion.tex
\section{Discussions and Lessons Learned}
\label{sec:discussion}

\begin{table*}
\centering
\caption{The responses for a subset of questions from the course evaluation questionnaire. 25 students answered the questionnaire using a Likert scale with 5 levels detailed below. For each question, the median answer is highlighted in bold and blue.}
\label{tab:course-evaluation}
\begin{tabularx}{\linewidth}{lXlllll}
\toprule
\textbf{ID} & \textbf{Question Description} & \textbf{1} & \textbf{2} & \textbf{3} & \textbf{4} & \textbf{5} \\
\midrule
\multicolumn{7}{c}{\textit{1:Never. 2:Rarely. 3:Sometimes. 4:Often. 5.Always}} \\
\midrule
Q1 & How often did you read the material before the lecture (Perusall or offline)? & 1 & 2 & 6 & \highcell{6} & 10 \\
Q2 & How often did you create or respond to annotations in Perusall? & 8 & \highcell{9} & 5 & 1 & 2  \\
Q3 & How often did you attend the lectures? & 1 & 0 & 0 & 2 & \highcell{22} \\
\midrule
\multicolumn{7}{c}{\textit{1:Strongly disagree. 2:Disagree. 3:Neutral. 4:Agree. 5:Strongly Agree}} \\
\midrule
Q4 & Reading the material before the lectures helped me better understand the lessons & 1 & 2 & 6 & \highcell{7} & 9  \\
Q5 & The lecture quizzes made me better understand the concepts being taught. & 2 & 2 & 4 & \highcell{10} & 6  \\
Q6 & Students are encouraged to ask questions and are given meaningful answers. & 0 & 0 & 1 & 9  & \highcell{15} \\
Q7 & I have learned something that I consider valuable. & 0 & 1 & 1 & 8 & \highcell{15} \\
Q8 & My interest in the subject has increased as a consequence of this course. & 1 & 1 & 2 & \highcell{9} & 12 \\
\bottomrule
\end{tabularx}
\end{table*}

Here, we complement our quantitative analysis above with a qualitative analysis of the course feedback provided by students. The course evaluation questionnaire reveals some qualitative aspects of the usage and the response of students to using Perusall as a tool. We also cover connections between Perusall and the student's learning and summarise our findings. We end the section with a summary of lessons learned and the limitations of our observations.

Course evaluations at our University are anonymous and use the Student Evaluation of Educational Quality (SEEQ) template~\cite{marsh1982seeq}. We extended the questionnaire with a few questions focusing on the social annotation aspect of the course. A subset of questions relevant to our discussion and their corresponding answers is presented in Table~\ref{tab:course-evaluation}. The response rate was 17\% (25 out of 142 students), which is low. One of the reasons for the low response rate is that the questionnaire was only available for students in the last 2-weeks of the course. Moreover, students reported that they received few reminders to complete the course evaluation.

Most students (64\%) would often or always read the material available in Perusall, but only three students stated that they create annotations in Perusall at the same frequency. 
Below, we share the statement from a student reporting that the need to annotate the material added a distraction to their studies, despite seeing their benefits when reading discussions from other students. Perusall has options to hide comments and annotations, but students did not receive a walk-through or demonstration of Perusall's features in course start. 

\interviewquote{Personally, I enjoyed being able to ask questions directly in Perusall and receiving answers. However, my personal learning style implies highlighting key concepts I find important, and sometimes in Perusall there were full paragraphs highlighted with a question, and it distracted me from the material}{Student}

After L12 (Inheritance), student representatives in the course asked the teacher to create anonymous annotations, which is an option for Perusall. The anonymity allows teachers to see the identity of students creating or replying to comments, but students do not see each other. We see a slight increase in activity from students after that, but this is still lower than some assignments before the anonymity was enabled.

The course evaluation questionnaire also includes two questions about the different teaching practices used in the course: (Q9) \textit{``What are the three things you liked the \textbf{least} about the course?''}, (Q10) \textit{``What are the three things you liked the \textbf{most} about the course?''}. Only one student listed Perusall and social annotation as one of the things they liked the least in the course. Particularly, the student was unsatisfied with the amount of time spent during the lecture quizzes (which typically cover the content from Perusall). Also, this student is more interested in a more traditional format of lectures where explanations are delivered predominantly by the lecturer rather than by their colleagues. Related work also reports that students struggle to adopt social annotations and move towards continuous learning~\cite{meyer2019SAinCS}.

\interviewquote{The discussions about the questions on Perusall. I think they were unnecessarily long and took away precious time from the lectures. I personally was interested in listening to the explanations from my lecturer, not from my peer that may be as lost as me.}{Student}

In contrast, a dominating number of students responded positively to the usage of social annotation and practical exercises during the lectures. From 18 text responses: 8 students (40\%) explicitly mentioned Perusall as one of the things they liked the most in the course, and 11 students (61\%) mentioned that the lecture quizzes and discussions helped them understand programming better. Particularly, students emphasised the scope, size and quality of the reading material created by the instructor for this course.

\interviewquote{The provided material on Perusall was on a nice level, and it was never unclear what to read before each lecture or where we were in the course. And lastly, the quizzes were a good measure of what had been understood or needed more practice, and also led to some nice discussions.}{Student}

Most of our analysis focuses on the correlation between Perusall annotations and exam points, such that we cannot use those measures to confidently infer causation between social annotation and students' learning. When analyzing the correlation between exam scores and students' learning, it is important to consider potential confounding factors. For instance, students who engage in social annotation might already be highly motivated and diligent, inherently contributing to their higher exam scores. Additionally, students vary in study habits or access to additional educational resources outside the course, which might influence exam scores independently of the course's teaching methods. Such an analysis would require other instruments, such as a thematic analysis of students' annotations throughout the course, as well as exercises or assessments that can show more of a progression throughout the course. We aim to perform those analyses in future work.

For the scope of this paper, we evaluate the learning based on the self-reported satisfaction from the course evaluation. The results suggest that social annotation correlates with the student's learning satisfaction. Note that 23 students (92\%) agree or strongly agree that they have learned something valuable in the course (Q7) and that, similarly, their interest (84\%) in programming increased as a consequence of the course (Q8). Therefore, we summarise our findings and lessons learned in the points below:

\begin{itemize}

\item Most students who engaged in social annotation passed the exam and, proportionally, showed higher grades. This is also reported in other areas such as physics~\cite{miller2018perusall} or multimedia applications~\cite{huang2008annotations}.

\item Completing more reading assignments in Perusall is correlated with higher passing rates.

\item Most students listed that Perusall, the lecture quizzes and class discussions among the three things they liked the most in the course.

\item More than 90\% of the course evaluation respondents agree that they learned something that they consider valuable, and their interest in programming has increased after the course.

\item Social annotation can leverage flipped classroom approaches. Many students read the material before the lecture and were motivated to engage in active learning during classes (e.g., Mentimeter quizzes and discussions).

\end{itemize}

Based on the experience reported in this paper, we make the following \textbf{recommendations} to instructors interested in introducing social annotation and Perusall to their programming courses:

\begin{itemize}
\item \textbf{Consider material length and scope:} Given that students in the analyzed course had approximately 24 hours between lectures to prepare and annotate the associated reading materials, it is crucial for these materials to be both concise and succinct. In this particular course, the average content for each lecture encompassed approximately 10 pages, comprising text and Java code examples

\item \textbf{Investigate incentives for social annotation:} More than half of the students consistently engaged with Perusall throughout the course. However, the percentage of completed reading assignments dropped over time. Increasing the number of bonus points, or making social annotation compulsory can encourage engagement.

\item \textbf{Demonstrate the social annotation platform early:} Students are not familiar with social annotation platforms in education, which can create initial barriers to engagement. Additionally, they may not be used to articulate their cognitive processes while writing their notes. To mitigate this, demonstrating Perusall, along with illustrative examples of both effective and ineffective annotations, can diminish the learning curve associated with social annotation.

\item \textbf{Explain the social annotation platform in the first weeks:} Students are not familiar with social annotation platforms, which can increase the friction of creating annotations in the first weeks of the course. Moreover, they are not necessarily critical about conveying their cognitive processes while studying. Showing how a tool like Perusall works, as well as some examples of meaningful (and not so meaningful) annotations can reduce the learning curve to social annotation as a practice.

\item \textbf{Enable anonymous annotations:} Initially, students may hesitate to openly express their uncertainties or questions to their peers. We observed a small increase in the number of annotations following the introduction of anonymous annotation, albeit introduced later in the course. It is worth considering that implementing anonymous annotations from the outset could have fostered greater engagement early on.
\end{itemize}

\subsection{Limitations}

There are some limitations in our analysis. One of the main construct validity threats is focusing on one instrument to indicate learning performance in the course. Written exams have limitations in conveying the learning of students due to various factors such as anxiety due to time constraints or cultural biases~\cite{nitko1996educational,grissom2015using}. On the other hand, exam results or grades provide a consistent way to compare trends across many instances of courses and have been used to evaluate teaching in software engineering education in the literature~\cite{gren2020flipped}. Using exam points also allowed us to compare our findings with other results from literature~ \cite{huang2008annotations,lin2013harnessing,miller2018perusall}. We mitigate the limitation in using exam scores by focusing our conclusions on the correlations between points, grades and Perusall activity without inferring a direct causation to learning.

Even though the instructor can override Perusall's automated grading, there are also risks with students receiving bonus points without making substantial comments (false positives). We did not make adjustments for those cases to avoid reducing the grade of the student. In our findings, most students did not earn bonus points as their comments were deemed insubstantial by Perusall. Less than five students requested a score review, and of those, just two had their scores modified. Nonetheless, we plan to explore Perusall's accuracy in future research further.

Moreover, course evaluation feedback is also subject to various factors, such as the student population (e.g., student bias), the impact of exam results, and the phrasing of the questions. We mitigate these factors by: (i) using the standardised SEEQ template~\cite{marsh1982seeq} and collecting course feedback before the written exam is performed. Moreover, course evaluations can be a useful tool to gather information about the student's experiences and can offer insights on how to improve course quality~\cite{johnson2007state}.

In our analysis, the heterogeneity of students' background knowledge in programming is a key internal validity threat. To assess this, we conducted an entry questionnaire at the course's outset, revealing that 67\% of students were unfamiliar with basic concepts like "conditionals." This suggests a relatively uniform knowledge level among participants. However, our results are not broadly generalizable due to the specific context of our study. Despite this, our large student sample and the range of topics align with those in many university programming courses. Future iterations of this course will incorporate student feedback to refine these teaching activities.

%% file: sections/06_conclusions.tex
\section{Conclusions and Future Work}
\label{sec:conclusion}

We report on our experience in introducing optional social annotation in a programming course via reading assignments using Perusall. We analyse whether social annotation has an impact on the student's grades and satisfaction. Our findings suggest that many (in our case, the majority of) students actively engage with the optional material and that a significant correlation to passing grades can be observed. However, only a subset of the annotations done by students are classified as meaningful. We also observe that many students interact with the material but leave assignments incomplete. Therefore, teachers aiming to use social annotation should account for the additional study time and effort required by students between lectures. Moreover, students might require guidance on how to use social annotation platforms and examples of how to create meaningful annotations.

Moreover, students were positive about the use of Perusall and, particularly, its combination with the quizzes and practical sessions during lectures. Students report that they felt that they learned something valuable in the course and that their interest in programming has increased. We argue that this latter aspect is particularly important, as an increased interest in programming can help students keep their motivation and engagement as they progress through different courses in the program. Although not the main focus of this study, the significant reduction in grading demands through the use of an AI grading tool indicates potential for scalability. With 140 students enrolled in the course and using Perusall, manually assessing each student's contributions would be impractical, even with numerous teaching assistants. We believe that teachers' time is better spent engaging in Perusall discussions, offering comments, and using these interactions to inform the preparation of exercises and practical sessions.

Future studies aim to compare our results to previous and future instances of the courses. Another important question is the relationship between social annotation and the other practical components of the course, such as the programming assignments done in groups or the lab sessions where students interact with teaching assistants. Lastly, we aim to compare our results to other topics in software engineering that require critical thinking and assessment in other complex constructs, such as software architectural design patterns, test specifications, or planning software development sessions. Most of these activities are typically carried out in teams and require consensus among participants, hence there might be an inherent component of collaborative learning in those activities that can be enhanced with tools such as Perusall.

%% file: article_acm.bbl

\begin{thebibliography}{29}


\ifx \showCODEN    \undefined \def \showCODEN     #1{\unskip}     \fi
\ifx \showDOI      \undefined \def \showDOI       #1{#1}\fi
\ifx \showISBNx    \undefined \def \showISBNx     #1{\unskip}     \fi
\ifx \showISBNxiii \undefined \def \showISBNxiii  #1{\unskip}     \fi
\ifx \showISSN     \undefined \def \showISSN      #1{\unskip}     \fi
\ifx \showLCCN     \undefined \def \showLCCN      #1{\unskip}     \fi
\ifx \shownote     \undefined \def \shownote      #1{#1}          \fi
\ifx \showarticletitle \undefined \def \showarticletitle #1{#1}   \fi
\ifx \showURL      \undefined \def \showURL       {\relax}        \fi
\providecommand\bibfield[2]{#2}
\providecommand\bibinfo[2]{#2}
\providecommand\natexlab[1]{#1}
\providecommand\showeprint[2][]{arXiv:#2}

\bibitem[Crouch and Mazur(2001)]%
        {crouch2001peer}
\bibfield{author}{\bibinfo{person}{Catherine~H Crouch} {and}
  \bibinfo{person}{Eric Mazur}.} \bibinfo{year}{2001}\natexlab{}.
\newblock \showarticletitle{Peer instruction: Ten years of experience and
  results}.
\newblock \bibinfo{journal}{\emph{American journal of physics}}
  \bibinfo{volume}{69}, \bibinfo{number}{9} (\bibinfo{year}{2001}),
  \bibinfo{pages}{970--977}.
\newblock


\bibitem[de~Oliveira~Neto and Dobslaw(2024)]%
        {replication2023perusall}
\bibfield{author}{\bibinfo{person}{Francisco~Gomes de Oliveira~Neto} {and}
  \bibinfo{person}{Felix Dobslaw}.} \bibinfo{year}{2024}\natexlab{}.
\newblock \bibinfo{booktitle}{\emph{{Analysis Package: Building Collaborative
  Learning: Exploring Social Annotation in Introductory Programming}}}.
\newblock
\urldef\tempurl%
\url{https://doi.org/10.5281/zenodo.10483184}
\showDOI{\tempurl}


\bibitem[Freeman et~al\mbox{.}(2014)]%
        {freeman2014active}
\bibfield{author}{\bibinfo{person}{Scott Freeman}, \bibinfo{person}{Sarah~L.
  Eddy}, \bibinfo{person}{Miles McDonough}, \bibinfo{person}{Michelle~K.
  Smith}, \bibinfo{person}{Nnadozie Okoroafor}, \bibinfo{person}{Hannah Jordt},
  {and} \bibinfo{person}{Mary~Pat Wenderoth}.} \bibinfo{year}{2014}\natexlab{}.
\newblock \showarticletitle{Active learning increases student performance in
  science, engineering, and mathematics}.
\newblock \bibinfo{journal}{\emph{Proceedings of the National Academy of
  Sciences}} \bibinfo{volume}{111}, \bibinfo{number}{23}
  (\bibinfo{year}{2014}), \bibinfo{pages}{8410--8415}.
\newblock
\urldef\tempurl%
\url{https://doi.org/10.1073/pnas.1319030111}
\showDOI{\tempurl}
\showeprint{https://www.pnas.org/doi/pdf/10.1073/pnas.1319030111}


\bibitem[Gao(2013)]%
        {gao2013case}
\bibfield{author}{\bibinfo{person}{Fei Gao}.} \bibinfo{year}{2013}\natexlab{}.
\newblock \showarticletitle{A case study of using a social annotation tool to
  support collaboratively learning}.
\newblock \bibinfo{journal}{\emph{The Internet and Higher Education}}
  \bibinfo{volume}{17} (\bibinfo{year}{2013}), \bibinfo{pages}{76--83}.
\newblock


\bibitem[Gren(2020)]%
        {gren2020flipped}
\bibfield{author}{\bibinfo{person}{Lucas Gren}.}
  \bibinfo{year}{2020}\natexlab{}.
\newblock \showarticletitle{A Flipped Classroom Approach to Teaching Empirical
  Software Engineering}.
\newblock \bibinfo{journal}{\emph{IEEE Transactions on Education}}
  \bibinfo{volume}{63}, \bibinfo{number}{3} (\bibinfo{year}{2020}),
  \bibinfo{pages}{155--163}.
\newblock
\urldef\tempurl%
\url{https://doi.org/10.1109/TE.2019.2960264}
\showDOI{\tempurl}


\bibitem[Grissom et~al\mbox{.}(2015)]%
        {grissom2015using}
\bibfield{author}{\bibinfo{person}{Jason~A Grissom}, \bibinfo{person}{Demetra
  Kalogrides}, {and} \bibinfo{person}{Susanna Loeb}.}
  \bibinfo{year}{2015}\natexlab{}.
\newblock \showarticletitle{Using student test scores to measure principal
  performance}.
\newblock \bibinfo{journal}{\emph{Educational evaluation and policy analysis}}
  \bibinfo{volume}{37}, \bibinfo{number}{1} (\bibinfo{year}{2015}),
  \bibinfo{pages}{3--28}.
\newblock


\bibitem[Huang et~al\mbox{.}(2008)]%
        {huang2008annotations}
\bibfield{author}{\bibinfo{person}{Yueh-Min Huang}, \bibinfo{person}{Tien-Chi
  Huang}, {and} \bibinfo{person}{Meng-Yeh Hsieh}.}
  \bibinfo{year}{2008}\natexlab{}.
\newblock \showarticletitle{Using annotation services in a ubiquitous Jigsaw
  cooperative learning environment}.
\newblock \bibinfo{journal}{\emph{Journal of Educational Technology \&
  Society}} \bibinfo{volume}{11}, \bibinfo{number}{2} (\bibinfo{year}{2008}),
  \bibinfo{pages}{3--15}.
\newblock
\showISSN{11763647, 14364522}
\urldef\tempurl%
\url{http://www.jstor.org/stable/jeductechsoci.11.2.3}
\showURL{%
\tempurl}


\bibitem[Hwang et~al\mbox{.}(2007)]%
        {hwang2007multimedia}
\bibfield{author}{\bibinfo{person}{Wu-Yuin Hwang}, \bibinfo{person}{Chin-Yu
  Wang}, {and} \bibinfo{person}{Mike Sharples}.}
  \bibinfo{year}{2007}\natexlab{}.
\newblock \showarticletitle{A study of multimedia annotation of Web-based
  materials}.
\newblock \bibinfo{journal}{\emph{Computers \& Education}}
  \bibinfo{volume}{48}, \bibinfo{number}{4} (\bibinfo{year}{2007}),
  \bibinfo{pages}{680--699}.
\newblock
\showISSN{0360-1315}
\urldef\tempurl%
\url{https://doi.org/10.1016/j.compedu.2005.04.020}
\showDOI{\tempurl}


\bibitem[Johnson et~al\mbox{.}(2007)]%
        {johnson2007state}
\bibfield{author}{\bibinfo{person}{David~W Johnson}, \bibinfo{person}{Roger~T
  Johnson}, {and} \bibinfo{person}{Karl Smith}.}
  \bibinfo{year}{2007}\natexlab{}.
\newblock \showarticletitle{The state of cooperative learning in postsecondary
  and professional settings}.
\newblock \bibinfo{journal}{\emph{Educational Psychology Review}}
  \bibinfo{volume}{19} (\bibinfo{year}{2007}), \bibinfo{pages}{15--29}.
\newblock


\bibitem[Kalir(2020)]%
        {kalir2020social}
\bibfield{author}{\bibinfo{person}{Jeremiah~H. Kalir}.}
  \bibinfo{year}{2020}\natexlab{}.
\newblock \showarticletitle{Social annotation enabling collaboration for open
  learning}.
\newblock \bibinfo{journal}{\emph{Distance Education}} \bibinfo{volume}{41},
  \bibinfo{number}{2} (\bibinfo{year}{2020}), \bibinfo{pages}{245--260}.
\newblock
\urldef\tempurl%
\url{https://doi.org/10.1080/01587919.2020.1757413}
\showDOI{\tempurl}


\bibitem[King et~al\mbox{.}(2019)]%
        {king2019instructional}
\bibfield{author}{\bibinfo{person}{Gary King}, \bibinfo{person}{Eric Mazur},
  \bibinfo{person}{Kelly Miller}, {and} \bibinfo{person}{Brian Lukoff}.}
  \bibinfo{year}{2019}\natexlab{}.
\newblock \bibinfo{title}{Instructional support platform for interactive
  learning environments}.
\newblock
\newblock
\newblock
\shownote{US Patent 10,438,498}.


\bibitem[Lee et~al\mbox{.}(2013)]%
        {lee2013can}
\bibfield{author}{\bibinfo{person}{Cynthia~Bailey Lee},
  \bibinfo{person}{Saturnino Garcia}, {and} \bibinfo{person}{Leo Porter}.}
  \bibinfo{year}{2013}\natexlab{}.
\newblock \showarticletitle{Can peer instruction be effective in upper-division
  computer science courses?}
\newblock \bibinfo{journal}{\emph{ACM Transactions on Computing Education
  (TOCE)}} \bibinfo{volume}{13}, \bibinfo{number}{3} (\bibinfo{year}{2013}),
  \bibinfo{pages}{1--22}.
\newblock


\bibitem[Lin and Lai(2013)]%
        {lin2013harnessing}
\bibfield{author}{\bibinfo{person}{Jian-Wei Lin} {and}
  \bibinfo{person}{Yuan-Cheng Lai}.} \bibinfo{year}{2013}\natexlab{}.
\newblock \showarticletitle{Harnessing Collaborative Annotations on Online
  Formative Assessments}.
\newblock \bibinfo{journal}{\emph{Journal of Educational Technology \&
  Society}} \bibinfo{volume}{16}, \bibinfo{number}{1} (\bibinfo{year}{2013}),
  \bibinfo{pages}{263--274}.
\newblock
\showISSN{11763647, 14364522}
\urldef\tempurl%
\url{http://www.jstor.org/stable/jeductechsoci.16.1.263}
\showURL{%
\tempurl}


\bibitem[Loksa and Ko(2016)]%
        {loksa2016selfregulation}
\bibfield{author}{\bibinfo{person}{Dastyni Loksa} {and} \bibinfo{person}{Amy~J.
  Ko}.} \bibinfo{year}{2016}\natexlab{}.
\newblock \showarticletitle{The Role of Self-Regulation in Programming Problem
  Solving Process and Success}. In \bibinfo{booktitle}{\emph{Proceedings of the
  2016 ACM Conference on International Computing Education Research}}
  (Melbourne, VIC, Australia) \emph{(\bibinfo{series}{ICER '16})}.
  \bibinfo{publisher}{Association for Computing Machinery},
  \bibinfo{address}{New York, NY, USA}, \bibinfo{pages}{83–91}.
\newblock
\showISBNx{9781450344494}
\urldef\tempurl%
\url{https://doi.org/10.1145/2960310.2960334}
\showDOI{\tempurl}


\bibitem[Loksa et~al\mbox{.}(2022)]%
        {loksa2022metacognition}
\bibfield{author}{\bibinfo{person}{Dastyni Loksa}, \bibinfo{person}{Lauren
  Margulieux}, \bibinfo{person}{Brett~A. Becker}, \bibinfo{person}{Michelle
  Craig}, \bibinfo{person}{Paul Denny}, \bibinfo{person}{Raymond Pettit}, {and}
  \bibinfo{person}{James Prather}.} \bibinfo{year}{2022}\natexlab{}.
\newblock \showarticletitle{Metacognition and Self-Regulation in Programming
  Education: Theories and Exemplars of Use}.
\newblock \bibinfo{journal}{\emph{ACM Trans. Comput. Educ.}}
  \bibinfo{volume}{22}, \bibinfo{number}{4}, Article \bibinfo{articleno}{39}
  (\bibinfo{date}{sep} \bibinfo{year}{2022}), \bibinfo{numpages}{31}~pages.
\newblock
\urldef\tempurl%
\url{https://doi.org/10.1145/3487050}
\showDOI{\tempurl}


\bibitem[Lopez et~al\mbox{.}(2008)]%
        {lopez2008tracing}
\bibfield{author}{\bibinfo{person}{Mike Lopez}, \bibinfo{person}{Jacqueline
  Whalley}, \bibinfo{person}{Phil Robbins}, {and} \bibinfo{person}{Raymond
  Lister}.} \bibinfo{year}{2008}\natexlab{}.
\newblock \showarticletitle{Relationships between Reading, Tracing and Writing
  Skills in Introductory Programming}. In \bibinfo{booktitle}{\emph{Proceedings
  of the Fourth International Workshop on Computing Education Research}}
  (Sydney, Australia) \emph{(\bibinfo{series}{ICER '08})}.
  \bibinfo{publisher}{Association for Computing Machinery},
  \bibinfo{address}{New York, NY, USA}, \bibinfo{pages}{101–112}.
\newblock
\showISBNx{9781605582160}
\urldef\tempurl%
\url{https://doi.org/10.1145/1404520.1404531}
\showDOI{\tempurl}


\bibitem[Marsh(1982)]%
        {marsh1982seeq}
\bibfield{author}{\bibinfo{person}{Herbert~W Marsh}.}
  \bibinfo{year}{1982}\natexlab{}.
\newblock \showarticletitle{SEEQ: A reliable, valid and useful instrument for
  collecting students' evaluation of University teaching.}
\newblock \bibinfo{journal}{\emph{British journal of educational psychology}}
  \bibinfo{volume}{52}, \bibinfo{number}{1} (\bibinfo{year}{1982}),
  \bibinfo{pages}{77--95}.
\newblock


\bibitem[Mazur(1997)]%
        {mazur1997peer}
\bibfield{author}{\bibinfo{person}{E. Mazur}.} \bibinfo{year}{1997}\natexlab{}.
\newblock \bibinfo{booktitle}{\emph{Peer Instruction: Getting Students to Think
  in Class}}.
\newblock \bibinfo{publisher}{American Institute of Physics},
  \bibinfo{pages}{981{\textendash}988}.
\newblock
\urldef\tempurl%
\url{https://doi.org/10.1016/0378-4371(79)90165-1}
\showDOI{\tempurl}


\bibitem[Meyer and M\"{u}ller(2019)]%
        {meyer2019SAinCS}
\bibfield{author}{\bibinfo{person}{M. Meyer} {and} \bibinfo{person}{T.
  M\"{u}ller}.} \bibinfo{year}{2019}\natexlab{}.
\newblock \showarticletitle{If it were that easy: First experiences of
  introducing a social learning platform in an undergraduate CS course}. In
  \bibinfo{booktitle}{\emph{ICERI2019 Proceedings}} (Seville, Spain)
  \emph{(\bibinfo{series}{12th annual International Conference of Education,
  Research and Innovation})}. \bibinfo{publisher}{IATED},
  \bibinfo{pages}{10688--10697}.
\newblock
\showISBNx{978-84-09-14755-7}
\showISSN{2340-1095}
\urldef\tempurl%
\url{https://doi.org/10.21125/iceri.2019.2624}
\showDOI{\tempurl}


\bibitem[Miller et~al\mbox{.}(2018)]%
        {miller2018perusall}
\bibfield{author}{\bibinfo{person}{Kelly Miller}, \bibinfo{person}{Brian
  Lukoff}, \bibinfo{person}{Gary King}, {and} \bibinfo{person}{Eric Mazur}.}
  \bibinfo{year}{2018}\natexlab{}.
\newblock \showarticletitle{Use of a Social Annotation Platform for Pre-Class
  Reading Assignments in a Flipped Introductory Physics Class}.
\newblock \bibinfo{journal}{\emph{Frontiers in Education}}  \bibinfo{volume}{3}
  (\bibinfo{year}{2018}), \bibinfo{pages}{1--11}.
\newblock
\showISSN{2504-284X}
\urldef\tempurl%
\url{https://doi.org/10.3389/feduc.2018.00008}
\showDOI{\tempurl}


\bibitem[Nelson et~al\mbox{.}(2017)]%
        {nelson2017comprehension}
\bibfield{author}{\bibinfo{person}{Greg~L. Nelson}, \bibinfo{person}{Benjamin
  Xie}, {and} \bibinfo{person}{Amy~J. Ko}.} \bibinfo{year}{2017}\natexlab{}.
\newblock \showarticletitle{Comprehension First: Evaluating a Novel Pedagogy
  and Tutoring System for Program Tracing in CS1}. In
  \bibinfo{booktitle}{\emph{Proceedings of the 2017 ACM Conference on
  International Computing Education Research}} (Tacoma, Washington, USA)
  \emph{(\bibinfo{series}{ICER '17})}. \bibinfo{publisher}{Association for
  Computing Machinery}, \bibinfo{address}{New York, NY, USA},
  \bibinfo{pages}{2–11}.
\newblock
\showISBNx{9781450349680}
\urldef\tempurl%
\url{https://doi.org/10.1145/3105726.3106178}
\showDOI{\tempurl}


\bibitem[Nitko(1996)]%
        {nitko1996educational}
\bibfield{author}{\bibinfo{person}{Anthony~J. Nitko}.}
  \bibinfo{year}{1996}\natexlab{}.
\newblock \bibinfo{booktitle}{\emph{Educational assessment of students}}.
\newblock \bibinfo{publisher}{ERIC}, \bibinfo{address}{Prentice-Hall Order
  Processing Center, P.O. Box 11071, Des Moines, IA 50336-1071}.
\newblock


\bibitem[Novak et~al\mbox{.}(2012)]%
        {novak2012educational}
\bibfield{author}{\bibinfo{person}{Elena Novak}, \bibinfo{person}{Rim Razzouk},
  {and} \bibinfo{person}{Tristan~E Johnson}.} \bibinfo{year}{2012}\natexlab{}.
\newblock \showarticletitle{The educational use of social annotation tools in
  higher education: A literature review}.
\newblock \bibinfo{journal}{\emph{The Internet and Higher Education}}
  \bibinfo{volume}{15}, \bibinfo{number}{1} (\bibinfo{year}{2012}),
  \bibinfo{pages}{39--49}.
\newblock


\bibitem[Porter et~al\mbox{.}(2011)]%
        {porter2011peer}
\bibfield{author}{\bibinfo{person}{Leo Porter}, \bibinfo{person}{Cynthia
  Bailey~Lee}, \bibinfo{person}{Beth Simon}, {and} \bibinfo{person}{Daniel
  Zingaro}.} \bibinfo{year}{2011}\natexlab{}.
\newblock \showarticletitle{Peer Instruction: Do Students Really Learn from
  Peer Discussion in Computing?}. In \bibinfo{booktitle}{\emph{Proceedings of
  the Seventh International Workshop on Computing Education Research}}
  (Providence, Rhode Island, USA) \emph{(\bibinfo{series}{ICER '11})}.
  \bibinfo{publisher}{Association for Computing Machinery},
  \bibinfo{address}{New York, NY, USA}, \bibinfo{pages}{45–52}.
\newblock
\showISBNx{9781450308298}
\urldef\tempurl%
\url{https://doi.org/10.1145/2016911.2016923}
\showDOI{\tempurl}


\bibitem[Qian and Lehman(2017)]%
        {qian2017students}
\bibfield{author}{\bibinfo{person}{Yizhou Qian} {and} \bibinfo{person}{James
  Lehman}.} \bibinfo{year}{2017}\natexlab{}.
\newblock \showarticletitle{Students’ misconceptions and other difficulties
  in introductory programming: A literature review}.
\newblock \bibinfo{journal}{\emph{ACM Transactions on Computing Education
  (TOCE)}} \bibinfo{volume}{18}, \bibinfo{number}{1} (\bibinfo{year}{2017}),
  \bibinfo{pages}{1--24}.
\newblock


\bibitem[Saeli et~al\mbox{.}(2011)]%
        {saeli2011teaching}
\bibfield{author}{\bibinfo{person}{Mara Saeli}, \bibinfo{person}{Jacob
  Perrenet}, \bibinfo{person}{Wim~MG Jochems}, {and} \bibinfo{person}{Bert
  Zwaneveld}.} \bibinfo{year}{2011}\natexlab{}.
\newblock \showarticletitle{Teaching programming in Secondary school: A
  pedagogical content knowledge perspective.}
\newblock \bibinfo{journal}{\emph{Informatics in education}}
  \bibinfo{volume}{10}, \bibinfo{number}{1} (\bibinfo{year}{2011}),
  \bibinfo{pages}{73--88}.
\newblock


\bibitem[Suhre et~al\mbox{.}(2019)]%
        {suhre2019students}
\bibfield{author}{\bibinfo{person}{Cor Suhre}, \bibinfo{person}{Koos Winnips},
  \bibinfo{person}{Vincent Boer}, \bibinfo{person}{Pablo Valdivia}, {and}
  \bibinfo{person}{Hans Beldhuis}.} \bibinfo{year}{2019}\natexlab{}.
\newblock \showarticletitle{Students{\textquoteright} experiences with the use
  of a social annotation tool to improve learning in flipped classrooms}. In
  \bibinfo{booktitle}{\emph{Fifth International Conference on Higher Education
  Advances}}. \bibinfo{publisher}{UPV Press}, \bibinfo{address}{Valencia,
  Spain}.
\newblock
\urldef\tempurl%
\url{https://doi.org/10.4995/HEAD19.2019.9131}
\showDOI{\tempurl}


\bibitem[Whalley et~al\mbox{.}(2006)]%
        {whalley2006australasian}
\bibfield{author}{\bibinfo{person}{Jacqueline~L. Whalley},
  \bibinfo{person}{Raymond Lister}, \bibinfo{person}{Errol Thompson},
  \bibinfo{person}{Tony Clear}, \bibinfo{person}{Phil Robbins},
  \bibinfo{person}{P.~K.~Ajith Kumar}, {and} \bibinfo{person}{Christine
  Prasad}.} \bibinfo{year}{2006}\natexlab{}.
\newblock \showarticletitle{An Australasian Study of Reading and Comprehension
  Skills in Novice Programmers, Using the Bloom and SOLO Taxonomies}. In
  \bibinfo{booktitle}{\emph{Proceedings of the 8th Australasian Conference on
  Computing Education - Volume 52}} (Hobart, Australia)
  \emph{(\bibinfo{series}{ACE '06})}. \bibinfo{publisher}{Australian Computer
  Society, Inc.}, \bibinfo{address}{AUS}, \bibinfo{pages}{243–252}.
\newblock
\showISBNx{1920682341}


\bibitem[Xie et~al\mbox{.}(2019)]%
        {xie2019theory}
\bibfield{author}{\bibinfo{person}{Benjamin Xie}, \bibinfo{person}{Dastyni
  Loksa}, \bibinfo{person}{Greg~L Nelson}, \bibinfo{person}{Matthew~J
  Davidson}, \bibinfo{person}{Dongsheng Dong}, \bibinfo{person}{Harrison Kwik},
  \bibinfo{person}{Alex~Hui Tan}, \bibinfo{person}{Leanne Hwa},
  \bibinfo{person}{Min Li}, {and} \bibinfo{person}{Amy~J Ko}.}
  \bibinfo{year}{2019}\natexlab{}.
\newblock \showarticletitle{A theory of instruction for introductory
  programming skills}.
\newblock \bibinfo{journal}{\emph{Computer Science Education}}
  \bibinfo{volume}{29}, \bibinfo{number}{2-3} (\bibinfo{year}{2019}),
  \bibinfo{pages}{205--253}.
\newblock


\end{thebibliography}
